\documentclass[reprint,superscriptaddress,showpacs,nofootinbib,amsmath,amssymb,aps,prl]{revtex4-1}
\usepackage{graphicx}
\usepackage{dcolumn}
\usepackage{bm}
\usepackage[utf8]{inputenc}
\usepackage[frenchb]{babel}
\usepackage{braket}

\begin{document}

\preprint{APS/123-QED}

\title{Self-localization of magnons and magnetoroton in a binary Bose-Einstein condensate}

\author{S. V. Andreev}
\email[Electronic adress: ]{Serguey.Andreev@gmail.com}
\affiliation{ITMO University, St. Petersburg 197101, Russia}
\author{O. I. Utesov}
\affiliation{National Research Center "Kurchatov Institute" B.P.\ Konstantinov Petersburg Nuclear Physics Institute, Gatchina 188300, Russia}
\affiliation{St. Petersburg Academic University - Nanotechnology Research and Education Centre of the Russian Academy of Sciences, 194021 St. Petersburg, Russia}
\affiliation{St. Petersburg State University, 7/9 Universitetskaya nab., St. Petersburg 199034, Russia}

\date{\today}

\begin{abstract}  
We consider a two-component Bose-condensed mixture characterized by positive $s$-wave scattering lengths. We assume equal densities and intra-species interactions. By doing the Bogoliubov transformation of an effective Hamiltonian we obtain the lower energy magnon dispersion incorporating the superfluid entrainment between the components. We argue that p-wave pairing of distinct bosons should be accompanied by self-localization of magnons and formation of a \textit{magnetoroton}. We demonstrate the effect on a model system of particles interacting via step potentials.
\end{abstract}


\maketitle

A roton, as initially postulated by Landau \cite{Landau1947}, is a dip in the energy dispersion of small-amplitude oscillations of the superfluid density. Rotons are ubiquitous in quantum liquids being close to solidification \cite{FeynmanIII, Beliaev, Brazovskii, Nozieres, Astrakharchik}. In gaseous Bose-Einstein condensates of atoms and excitons in semiconductors the roton may be mimicked by softening the contact part of the interaction and adding long-range dipolar repulsion \cite{Santos1, Zillich, Lu, Andreev1}. Upon compression the dipolar condensate was predicted to transform into the supersolid: a lattice with macroscopically populated sites which combines the properties of a crystal and a superfluid \cite{Lu, Andreev1}. Quest for this exotic state of matter has been at the frontier of the quantum physics over the past years \cite{Ferlaino, Pfau, Tanzi}.

Even richer phenomenology is expected for binary Bose mixtures. In addition to the familiar phonon mode, the spectrum of a mixture is known to contain the magnon mode, which corresponds to the out-of-phase fluctuations of the components \cite{Pitaevskii, Pethick}:
\begin{equation}
\label{spin0}
\varepsilon_\mathrm{m}(\bm p)=\sqrt{\frac{\hbar^2 p^2}{2m}\left(\frac{\hbar^2 p^2}{2m}+2n[g_{\uparrow\uparrow}(\bm p)-g_{\uparrow\downarrow}(\bm p)]\right)},
\end{equation}
where we have defined $g_{\sigma\sigma^\prime}(\bm p)\equiv-4\pi\hbar^2/m f_{\sigma\sigma^\prime}(\bm p)$ with $f_{\sigma\sigma^\prime}(\bm p)$ being the (on-shell) scattering amplitudes of two particles in vacuum. Here, and in what follows, we assume positive $s$-wave scattering lengths $a_{\sigma\sigma^\prime}=-f_{\sigma\sigma^\prime}(0)>0$ and identical intra-species interactions, $g_{\uparrow\uparrow}(\bm p)\equiv g_{\downarrow\downarrow}(\bm p)$, which implies equal densities of the components  $n_{\uparrow}=n_{\downarrow}\equiv n$ in the thermodynamic equilibrium. The free-particle form of Eq. \eqref{spin0} at $a_{\uparrow\uparrow}=a_{\uparrow\downarrow}$ corresponds to the vanishing spin-wave velocity $c_s=\sqrt{4\pi n(a_{\uparrow\uparrow}-a_{\uparrow\downarrow})}\hbar/m$ and reflects the analogy of the SU(2)-symmetric system to a Heisenberg ferromagnet \cite{Halperin1}. For $g_{\uparrow\uparrow}(\bm p)< g_{\uparrow\downarrow}(\bm p )$ the components tend to spatially separate. This so-called immiscibility transition may occur either at $\bm p=0$ \cite{Colson, Ho, Jiang} or at $\bm p\neq 0$ \cite{Wilson}. Similar to the roton softening of the density mode, the latter case requires increasing $n$ above some critical value. Note, that the $\bm p=0$ transition may also show periodic textures due to faster growth of instabilities with wavevectors $p\sim m \lvert c_s\rvert/\hbar$ \cite{Timmermans}. Whereas such transient phenomena can be observed in condensates with contact interactions \cite{Sadler, SpinorBEC}, realization of a stable "roton immiscibility" would require momentum-dependent pseudo-potentials. As in the case of supersolids, the incumbent candidates are dipolar species \cite{FerlainoMixtures, SpinorBEC}.

The important new ingredient in the theoretical description of Bose mixtures as compared to scalar condensates is the $p$-wave scattering of distinct species. As we show in this Letter, the $p$-waves are not fully accounted for by the textbook formula \eqref{spin0}. As a result, the physics of magnons outlined above misses the polaronic effect, which was recently shown \cite{Utesov} to be responsible for the superfluid entrainment (``quantum friction") between the components \cite{AB}. The effect becomes significant on approaching a $p$-wave resonance from the attractive side. By using a properly generalized Bogoliubov transformation we obtain at $\bm p\rightarrow 0$ 
\begin{equation}
\label{spinApprox}
\varepsilon_\mathrm{m}^\ast(\bm p)=\sqrt{\frac{\hbar^2 p^2}{2m_\ast}\left(\frac{\hbar^2 p^2}{2m_\ast}+2n[g_{\uparrow\uparrow}(0)-g_{\uparrow\downarrow}(0)]\right)},
\end{equation}
where
\begin{equation}
\label{mass}
 m_\ast=\frac{m}{1-12\pi n\lvert\upsilon\rvert},
\end{equation}
with $\upsilon <0$ being the $p$-wave scattering volume. The magnon drags surrounding particles, which increases its effective mass. As $n$ approaches the critical value $n_c^{(1)}=(12\pi \lvert\upsilon\rvert)^{-1}$, the mixture becomes dynamically unstable. The result \eqref{mass} thus suggests a new mechanism for the phase separation: \textit{self-localization} of magnons. The self-localization may also occur at a finite momentum: the bare magnons with energies closer to the resonance exhibit stronger entrainment and may achieve zero group velocity at $n_c^{(2)}<n_c^{(1)}$. We call the corresponding dip in the dispersion curve \textit{magnetoroton}. In contrast to the roton immiscibility, the magnetoroton does not require going beyond the contact interactions. Another compelling property is that in addition to the polarization, the new quasiparticle carries also an angular momentum. A hypothetical self-localized \textit{magnon crystal} might compete with (if preclude) the finite-momentum atomic-molecular and spinor molecular superfluids \cite{RPsuperfluids}. We illustrate the self-localization of magnons on a model system with step two-body potentials. Step-like potential is a simple and yet insightful approximation widely employed in studies of the roton phenomena \cite{Rica, Henkel, SoftSpheres, Pines2}. The model allows analytical mean-field description of all relevant regimes: the Landau roton in the density mode, roton immiscibility and the self-localized magnetoroton.   

We first give the derivation of the general result \eqref{spinApprox} and discuss the underlying physics in more detail. The second-quantized Hamiltonian of the system reads
\begin{equation}
\label{Hamiltonian}
\begin{split}
\hat H&=\sum_{\textbf{p},\sigma}\frac{\hbar^2p^2}{2m}\hat a_{\sigma, \textbf{p}}^{\dag} \hat a_{\sigma, \textbf{p}}+\\
&\frac{1}{2V}\sum_{\textbf p_1,\textbf p_2,\textbf{q},\sigma,\sigma^\prime}\hat a_{\sigma, \textbf p_1+\textbf q}^{\dag} \hat a_{\sigma^\prime,\textbf p_2-\textbf q}^{\dag} V_{\sigma\sigma^\prime}(\textbf{q})\hat a_{\sigma, \textbf p_1}\hat a_{\sigma^\prime,\textbf p_2}.
\end{split}
\end{equation}
Here $V_{\sigma\sigma^\prime}(\textbf{q})$ are the Fourier transforms of the microscopic two-body potentials and the boson operators $\hat a_{\sigma, \bm p}$ obey the commutation relations
\begin{equation}
[\hat a_{\sigma, \bm p_1},\hat a_{\sigma', \bm p_2}^\dagger]=\delta_{\sigma\sigma',\bm p_1\bm p_2}.
\end{equation}  
To study the low-energy properties of the model \eqref{Hamiltonian} we employ the effective Hamiltonian
\begin{widetext}
\begin{equation}
\label{EffectiveHamiltonian}
\hat H_\ast=\sum_{\textbf{p},\sigma}\frac{\hbar^2 p^2}{2m}\hat a_{\sigma, \textbf{p}}^{\dag} \hat a_{\sigma, \textbf{p}}+\frac{1}{2V}\sum_{\textbf k,\textbf p,\textbf q,\sigma,\sigma^\prime}\hat a_{\sigma, \textbf k+\textbf p}^{\dag} \hat a_{\sigma^\prime,\textbf k-\textbf p}^{\dag} g_{\sigma\sigma^\prime}(\bm p, \bm q)\hat a_{\sigma, \bm k+\bm q}\hat a_{\sigma^\prime,\bm k-\bm q},
\end{equation}
\end{widetext}
where we have defined $g_{\sigma\sigma^\prime}(\bm p, \bm q)\equiv-4\pi\hbar^2/m f_{\sigma\sigma^\prime}(\bm p, \bm q)$ with $f_{\sigma\sigma^\prime}(\bm p, \bm q)$ being the off-shell scattering amplitudes of two particles in vacuum. In the limit of weak interactions the Hamiltonian \eqref{EffectiveHamiltonian} correctly reproduces the collective behaviour of a mixture revealed by the general diagrammatic theory \cite{Utesov}. To zero order one may replace $\hat a_{\sigma, 0}$ by the $c$-numbers $\sqrt{N_\uparrow}=\sqrt{N_\downarrow}=\sqrt{N}$ and find $E_0=1/2nN[g_{\uparrow\uparrow}(0,0)+g_{\uparrow\downarrow}(0,0)]$ and $\mu=n[g_{\uparrow\uparrow}(0,0)+g_{\uparrow\downarrow}(0,0)]$ for the condensate energy and chemical potential, respectively.  By retaining the quadratic terms in the operators $\hat a_{\sigma, \bm p}$ with $\bm p\neq 0$ and applying the Bogoliubov transformation, one obtains the elementary excitation spectrum, the major focus of this paper:
\begin{widetext}
\begin{subequations}
\label{Spectrum}
\begin{align}
\label{Phonon}
\varepsilon_\mathrm{ph}^\ast(\bm p)&=\sqrt{\left(\frac{\hbar^2 p^2}{2m}+n\left[2g_{\uparrow\uparrow}^{+}\left(\tfrac{\bm p}{2},\tfrac{\bm p}{2}\right)+2g_{\uparrow\downarrow}^{+}\left(\tfrac{\bm p}{2},\tfrac{\bm p}{2}\right)-g_{\uparrow\uparrow}(0,0)-g_{\uparrow\downarrow}(0,0)\right]\right)^2-n^2\left[g_{\uparrow\uparrow}(0,\bm p)+g_{\uparrow\downarrow}(0,\bm p)\right]^2}\\
\label{Spin}
\varepsilon_\mathrm{m}^\ast(\bm p)&=\sqrt{\left(\frac{\hbar^2 p^2}{2m}+n\left[2g_{\uparrow\uparrow}^{+}\left(\tfrac{\bm p}{2},\tfrac{\bm p}{2}\right)+2g_{\uparrow\downarrow}^{-}\left(\tfrac{\bm p}{2},\tfrac{\bm p}{2}\right)-g_{\uparrow\uparrow}(0,0)-g_{\uparrow\downarrow}(0,0)\right]\right)^2-n^2\left[g_{\uparrow\uparrow}(0,\bm p)-g_{\uparrow\downarrow}(0,\bm p)\right]^2},
\end{align}
\end{subequations}
\end{widetext}
where
\begin{equation}
g_{\sigma\sigma'}^{\pm}(\bm p,\bm p)=\tfrac{1}{2}[g_{\sigma\sigma'}(\bm p,\bm p)\pm g_{\sigma\sigma'}(\bm p,-\bm p)].
\end{equation}  

For identical intra- and inter-component interactions Eq. \eqref{Phonon} turns into the Beliaev result for the phonon mode of a scalar condensate \cite{Beliaev}
\begin{equation}
\label{BeliaevPhonon}
\begin{split}
&\varepsilon_\mathrm{ph}^\ast(\bm p)=\\
&\sqrt{\left[\frac{\hbar^2 p^2}{2m}+4ng^{+}\left(\tfrac{\bm p}{2},\tfrac{\bm p}{2}\right)-2ng(0,0)\right]^2-4n^2g^2(0,\bm p)},
\end{split}
\end{equation}
whereas the lower branch \eqref{Spin} takes the "ferromagnetic" form
\begin{equation}
\label{Magnon}
\varepsilon_\mathrm{m}^\ast(\bm p)=\frac{\hbar^2 p^2}{2m}+2n[g\left(\tfrac{\bm p}{2},\tfrac{\bm p}{2}\right)-g(0,0)].
\end{equation}
The proper combination of the effective potentials $g(\bm p,\bm q)$ in \eqref{BeliaevPhonon} captures the short-range correlations which build up the Landau roton in the strongly-correlated regime \cite{Beliaev, Ronen, Zwerger}. As the gas parameter $na^3$ approaches the unity, the phonon dispersion first develops an inflection and finally a maximum followed by a minimum. According to Feynman \cite{FeynmanIII}, the roton wavefunction in the configuration space may be compared to that of a slow impurity particle pushing through the dense media. One may notice, that it bears a resemblance to the Landau-Pekar ansatz for a self-localized electron in a polar crystal \cite{Pekar}. Their description was subsequently adopted to a Bose polaron in the strongly-coupled regime \cite{Timmermans1}. It is thus tempting to associate crystallization of a superfluid with self-localization of rotons.

Although some signatures of strong correlations in the spectrum of a dilute BEC have been detected \cite{Experiments}, the appealing scenario postulated above is likely to be hindered by quantum fluctuations. Smallness of the quantum depletion of the condensate requires $na^3\ll1$. The leading correction to the quasiparticle mass in this limit comes from the second order of the Bogoliubov approach due to interaction of the roton with a virtual cloud of phonons. In the Feynman language this corresponds to adding a "backflow" of particles analogous to a vortex ring \cite{Feynman2, Miller, Pines1}. The correction is directly proportional to the quantum depletion of the condensate. 

The situation is very different for the magnon branch \eqref{Magnon}. Here, not only the $s$-waves, but also $p$-waves contribute to the renormalization of the quasi-particle mass. This offers a possibility to explore strong polaronic effects at small $na^3$ by means of an interspecies $p$-wave Feshbach resonance. Close to the resonance the second term in Eq. \eqref{Magnon} is governed by the effective range expansion for the $p$-wave scattering amplitude \cite{Taylor}:
\begin{equation}
\label{fp}
f_p(k)=\frac{k^2}{-\upsilon^{-1}+k_0 k^2/2-ik^3}.
\end{equation}
The $p$-wave scattering volume scales as $\upsilon\sim -1/\nu$, where $\nu$ is the detuning. We are only concerned with $\nu>0$, since at $\nu<0$ a thermodynamically stable many-body phase would be a spinor molecular superfluid \cite{RPsuperfluids}. Assuming $\lvert \upsilon \rvert\gg a^3$ and taking $f_p(k)=-\upsilon k^2$ at $k\rightarrow 0$ one obtains
\begin{equation}
\label{parabola}
\varepsilon_\mathrm{m}^\ast(\bm p)=\frac{\hbar^2 p^2}{2m_\ast},
\end{equation}
where $m_\ast$ is given by Eq. \eqref{mass}. This is in stark contrast with what one would expect on the basis of Eq. \eqref{spin0}: the latter predicts the free-particle dispersion with the bare mass $m$. The magnon core can also be dressed by "backflow" currents \cite{Utesov, Bruun1}, but the corresponding correction would be negligible in the typical experimental conditions \cite{Kim}.        

\begin{figure}[t]
\label{Fig1}

\centering

  \begin{tabular}{c}


    \includegraphics[width=0.9\columnwidth]{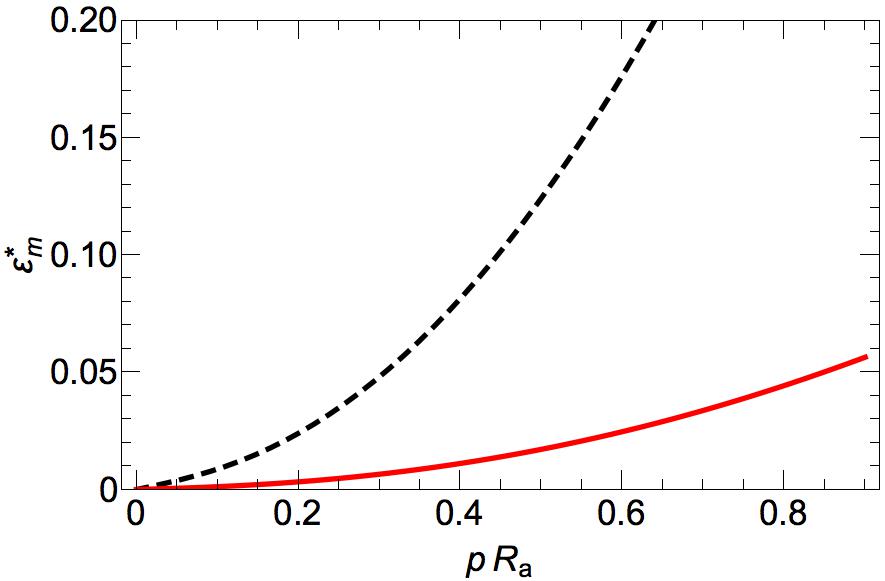}\\

    \includegraphics[width=0.9\columnwidth]{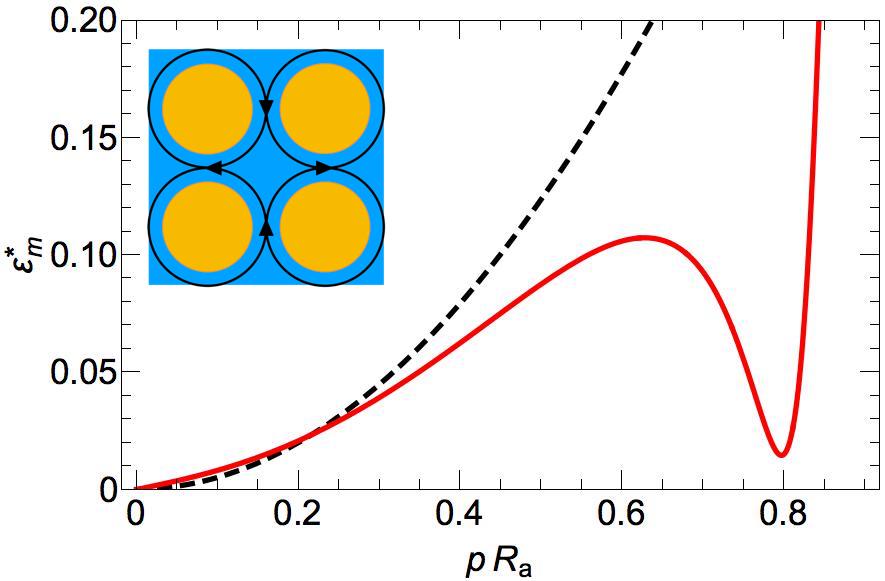}\\

  \end{tabular}
  
\caption{The magnon dispersion \eqref{Spin} calculated for the step interaction potentials \eqref{potentials}. The inter-component interaction contains a $p$-wave resonance. In stark contrast with the prediction of Eq. \eqref{spin0}, the spin-wave velocity calculated by using the formula \eqref{Spin} vanishes, as the density approaches the critical value defined by Eq. \eqref{mass} (upper panel). The values of parameters are $\alpha=2.7$, $\beta=2.7$, $\sigma=0.5$, $nR_a^3=0.01$ (dashed line) and $nR_a^3=0.1746$ (solid line). At larger values of the $p$-wave scattering volume [smaller detuning \eqref{nu}] the dispersion develops a minimum (lower panel). Here $\alpha=3$, $\beta=3.12$, $nR_a^3=0.0001$ (dashed line) and $nR_a^3=0.0033$ (solid line). The inset shows a unit cell of the hypothetical magnon crystal (in the cross-section). The two components occupy two different regions in space: inner circles and the surrounding area. Surrounding circles with arrows indicate directions of spontaneous currents.}  

\end{figure}    

Along the same lines, one may derive Eq. \eqref{spinApprox} from \eqref{Spin}. Increasing the density results in dramatic enhancement of the magnon mass, so that the spin-wave velocity $c_s^\ast=\sqrt{4\pi n(a_{\uparrow\uparrow}-a_{\uparrow\downarrow})}\hbar/m_\ast$ may approach zero at $a_{\uparrow\uparrow}>a_{\uparrow\downarrow}$, i.e. in the region of the standard miscibility defined by Eq. \eqref{spin0}. Basing on the results of Ref. \cite{Timmermans2} we may postulate that self-localization of magnons represent the nucleation process for a new phase separation transition. At sufficiently large $\lvert \upsilon \rvert$ (small $\nu$) this transition may occur at $\bm p\neq 0$:  the magnons with energies closer to the pole of the scattering amplitude \eqref{fp} exhibit stronger entrainment due to pairing of distinct particles. As one increases the density above $n_c^{(2)}\propto \nu/\lvert \upsilon \rvert$, the dispersion \eqref{Spin} develops a minimum, which eventually touches zero. In the relevant limit $\upsilon k_0^3\gg 1$ the position of the magnetoroton scales as
\begin{equation}
\label{pmr}
p_r\sim \sqrt{\nu}.
\end{equation}
In contrast to the Landau-Feynman rotons, the core of the magnetoroton should posses an angular momentum. Pairing of the components is known to result in a spinor molecular superfluidity when crossing the resonance \cite{RPsuperfluids}. The phase separation precludes formation of the molecular condensate. Instead, one may imagine a lattice of one component surrounded by circulating currents of the other. Long-wavelength structures could correspond to formation of bi-rotons and larger roton complexes. A unit cell of the \textit{magnon crystal} is schematically illustrated in the inset of Fig. 1. The direction of the current flow alternates, so that the total angular momentum of the system remains equal to zero.

To illustrate the general arguments given above, we take the model two-body potentials
\begin{equation}
\label{potentials}
V_{\sigma\sigma'}(\bm r)=\begin{cases}
U_{\sigma\sigma'},&r\leqslant R_{\sigma\sigma'}\\
0,&r> R_{\sigma\sigma'},
\end{cases}
\end{equation}
where we let $U_{\uparrow\uparrow}=U_{\downarrow\downarrow}\equiv U_a$, $U_{\uparrow\downarrow}\equiv U_b$, $R_{\uparrow\uparrow}=R_{\downarrow\downarrow}\equiv R_a$ and $R_{\uparrow\downarrow}\equiv R_b$. The useful dimensionless combinations of parameters are $\alpha= \sqrt{mU_aR_a^2/\hbar^2}$, $\beta=\sqrt{ m\lvert U_b\rvert R_b^2/\hbar^2}$ and $\sigma=R_b/R_a$. For one-component condensates step potentials have been extensively used to simulate the roton phenomena in weakly-interacting systems \cite{Rica, Henkel}. The radius $R$ of the potential may greatly exceed the average inter-particle distance $n^{-1/3}$ at the expense of the potential barrier $U$, which must be small in order to fulfil the criterion of weak interactions $n a^3\ll 1$, where $a\sim\alpha R$. At $k\sim R^{-1}$ the scattering amplitude takes negative values which favours crystallization of the condensate density. Such "droplet crystals" with supersolid properties have indeed been observed in numerical simulations in 3D, 2D and 1D geometries \cite{Henkel, SoftSpheres, Andreev2}.

For two-component mixtures the limit of large $nR_{\sigma\sigma'}^3$ and small $U_{\sigma\sigma'}$ also contains the roton immiscibility previously predicted for dipolar systems \cite{Wilson}. Our calculations show, that the effect of entrainment is negligible in this case. Various contributions in \eqref{Spin} at $p\rightarrow 0$ cancel each other and we find $m_\ast=m$, so that one may safely use the familiar expression \eqref{spin0} [where $g_{\sigma\sigma'}(\bm p)$ can be taken in the Born approximation]. At $R_b<R_a$ approaching the standard $\bm p=0$ phase separation transition and increasing the density results in appearance of a roton-like minimum at
 \begin{equation}
 \label{kR} 
 p_i\sim \sqrt{\delta},
 \end{equation}
 where the dimensionless parameter
 \begin{equation}
 \label{miscibility}
\delta=1-\beta R_b/\alpha R_a
\end{equation}
shows how far we are from the miscibility boundary (defined by $\delta=0$). More detailed account for the roton immiscibility in a gas of soft spheres will be given in a separate paper. Here we only aim to demonstrate a profound difference of this phenomenon from the magnetoroton, unpredicted within the formula \eqref{spin0}.

The magnetoroton does not require long-range interactions, which in our toy model have been mimicked by letting $R_{\sigma\sigma'}$'s be macroscopically large. Instead, one needs to enhance the polaronic effect. This can be done by taking $U_b<0$ and choosing the parameter $\beta$ such that there is a bound state (and $a_{\uparrow\uparrow/\downarrow\downarrow}>a_{\uparrow\downarrow}>0$). The centrifugal barrier for the scattering of particles with $l=1$ transforms this state into a $p$-wave resonance with positive energy and finite lifetime. The resonance absorbs the $p$-wave scattering-state wave function into the core of the interaction potential, thus inducing strong quantum friction between the components. The strength of the effect is governed by the detuning \cite{SI}
\begin{equation}
\label{nu}
\nu=\pi^2-\beta^2
\end{equation}
according to the general consideration given below Eq. \eqref{fp}. In Fig. 1 we show first the situation, where self-localization occurs at $\bm p=0$ (upper panel), and then the case of the scattering volume $\lvert\upsilon\rvert$ being sufficiently large to feature the magnetoroton (lower panel). The position of the magnetoroton scales with the detuning [Eq. \eqref{pmr}] and, in contrast to the roton immiscibility [Eq. \eqref{kR}], is not sensitive to the proximity to the phase separation boundary. The new phenomena occur in the dilute regime: both Landau rotons in the density mode and roton immiscibility would require far larger values of $nR^3$.

Interspecies $p$-wave Feshbach resonances have been predicted for $^{85}$Rb-$^{87}$Rb mixture \cite{Ticknor}. Subsequent experiments on this system revealed formation of alternating domains of single-species condensates \cite{Papp}. Other possible settings to check include dipolar atoms, molecules \cite{Klawun} and excitons \cite{SantosExcitons} in bilayers. In the latter case the effects of the reduced dimensionality should be taken into account \cite{Utesov}.

Self-localized magnetorotons constitute a new class of quasiparticles which combine the properties of a polarization wave with those of a Bose polaron. Their behaviour is governed by $p$-wave interactions, which can be made strong without destroying the condensate. This is in contrast to the usual rotons in the strongly-correlated He II and single-species Bose-Einstein condensates close to unitarity. The magnetoroton carries an angular momentum, a fingeprint of the spinor molecular superfluidity \cite{RPsuperfluids, footnote}. Phase separation associated with the magnetoroton instability precludes formation of the paired condensate. Instead, the microscopic angular momenta may transform into circulating currents in a hypothetical magnon crystal. This new state of matter would furnish a spectacular visualization of a duality between the crystallization phenomena in quantum liquids and the physics of a Bose polaron.        

S. V. acknowledges the support by Russian Science Foundation (Grant No. 18-72-00013). Contribution to the study by O. Utesov was funded by RFBR according to the research project 18-02-00706.


\begin{thebibliography}{1}

\bibitem{Landau1947} L. D. Landau, J. Phys. USSR \textbf{11}, 91 (1947).

\bibitem{FeynmanIII} R. P. Feynman, Phys. Rev. \textbf{94}, 262 (1954).

\bibitem{Beliaev} S. T. Beliaev, Zh. Eksp. Teor. Fiz. \textbf{34}, 417 (1958) [Sov. Phys. JETP \textbf{7}, 289 (1958)].

\bibitem{Brazovskii} S. A. Brazovskii, Sov. Phys. JETP \textbf{41}, 85 (1975).

\bibitem{Nozieres} P. Nozieres, J. Low Temp. Phys. \textbf{137}, 45 (2004).

\bibitem{Astrakharchik} G. E. Astrakharchik, J. Boronat, I. L. Kurbakov, and Yu. E. Lozovik, Phys. Rev. Lett. \textbf{98}, 060405 (2007).

\bibitem{Santos1} L. Santos, G. V. Shlyapnikov, and M. Lewenstein, Phys. Rev. Lett. \textbf{90}, 250403 (2003); D. H. J. O Dell, S. Giovanazzi, and G. Kurizki
Phys. Rev. Lett. \textbf{90}, 110402 (2003).

\bibitem{Zillich} A. Macia, D. Hufnagl, F. Mazzanti, J. Boronat, and R. E. Zillich
Phys. Rev. Lett. \textbf{109}, 235307 (2012).

\bibitem{Lu} Z.-K. Lu, Y. Li, D. S. Petrov, and G. V. Shlyapnikov, Phys. Rev. Lett. \textbf{115}, 075303 (2015).

\bibitem{Andreev1} S. V. Andreev, Phys. Rev. B \textbf{92}, 041117(R) (2015); S. V. Andreev, Phys. Rev. B \textbf{94}, 140501(R) (2016); S. V. Andreev, Phys. Rev. B \textbf{95}, 184519 (2017).

\bibitem{Pfau} H. Kadau, M. Schmitt, M. Wenzel, C. Wink, T. Maier, I. Ferrier-Barbut, and T. Pfau, Nature (London) \textbf{530}, 194 (2016); F. Bottcher, J.-N. Schmidt, M. Wenzel, J. Hertkorn, M. Guo, T. Langen, and T. Pfau, Phys. Rev. X \textbf{9}, 011051 (2019).

\bibitem{Ferlaino} L. Chomaz, R. M. W. van Bijnen, D. Petter, G. Faraoni, S. Baier, J. H. Becher, M. J. Mark, F. Wachtler, L. Santos, F. Ferlaino, Nature Physics (2018); L. Chomaz et al., Phys. Rev. X \textbf{9}, 021012 (2019).

\bibitem{Tanzi} L. Tanzi, E. Lucioni, F. Fama, J. Catani, A. Fioretti, C. Gabbanini, R. N. Bisset, L. Santos, and G. Modugno, Phys. Rev. Lett. \textbf{122}, 130405 (2019).

\bibitem{Pitaevskii} L. P. Pitaevskii and S. Stringari, \textit{Bose-Einstein Condensation and Superfluidity} (Oxford University Press, Oxford,
2016).

\bibitem{Pethick} C. J. Pethick and H. Smith, \textit{Bose-Einstein Condensation in Dilute Bose Gases} (Cambridge University Press, Cambridge, 2008).

\bibitem{Halperin1} B. I. Halperin, Phys. Rev. B \textbf{11}, 178 (1975).

\bibitem{Colson} W. B. Colson and A. L. Fetter, J. Low Temp. Phys. \textbf{33}, 231 (1978).

\bibitem{Ho} Tin-Lun Ho and V. B. Shenoy, Phys. Rev. Lett. \textbf{77}, 3276 (1996).

\bibitem{Jiang} Xunda Jiang et al., New J. Phys. \textbf{21}, 023014 (2019).

\bibitem{Wilson} R. M. Wilson, C. Ticknor, J. L. Bohn, and E. Timmermans, Phys. Rev. A \textbf{86}, 033606 (2012).

\bibitem{Timmermans} E. Timmermans, Phys. Rev. Lett. \textbf{81}, 5718 (1998).

\bibitem{Sadler} L. Sadler, J. M. Higbie, S. R. Leslie, M. Vengalattore and D. M. Stamper-Kurn, Nature (London) \textbf{443}, 312 (2006).

\bibitem{SpinorBEC} See, however, the subsequent papers M. Vengalattore, S. R. Leslie, J. Guzman, and D. M. Stamper-Kurn
Phys. Rev. Lett. 100, 170403 (2008); Yujiro Eto, Hiroki Saito, and Takuya Hirano, Phys. Rev. Lett. 112, 185301 (2014); S. Lepoutre, L. Gabardos, K. Kechadi, P. Pedri, O. Gorceix, E. Marechal, L. Vernac, and B. Laburthe-Tolra, Phys. Rev. Lett. 121, 013201 (2018) showing the influence of the magnetic dipolar forces on the collective behaviour of spinor condensates.

\bibitem{FerlainoMixtures} A. Trautmann, P. Ilzhofer, G. Durastante, C. Politi, M. Sohmen, M. J. Mark, and F. Ferlaino, Phys. Rev. Lett. \textbf{121}, 213601 (2018).

\bibitem{Utesov} O. I. Utesov, M. I. Baglay and S. V. Andreev, Phys. Rev. A \textbf{97}, 053617 (2018).

\bibitem{AB} A. F. Andreev and E. P. Bashkin, Sov. Phys. JETP \textbf{42}, 164 (1976).

\bibitem{RPsuperfluids} A. B. Kuklov, Phys. Rev. Lett. \textbf{97}, 110405 (2006); W. V. Liu and C. Wu, Phys. Rev. A \textbf{74}, 013607 (2006); L. Radzihovsky and S. Choi, Phys. Rev. Lett. \textbf{103}, 095302
(2009); S. Choi and L. Radzihovsky, Phys. Rev. A \textbf{84}, 043612 (2011).     

\bibitem{Pines2} C. H. Aldrich and D. Pines. J . Low Temp. Phys. \textbf{25}, 677 (1976); \textbf{32}, 689 (1978); D. Pines, Can. Journ. Phys. \textbf{65}, 1357 (1987). 

\bibitem{Rica} Yves Pomeau and Sergio Rica, Phys. Rev. Lett. \textbf{71}, 247 (1993); Yves Pomeau and Sergio Rica
Phys. Rev. Lett. 72, 2426 (1994); N. Henkel, R. Nath and T. Pohl, Phys. Rev. Lett. \textbf{104}, 195302 (2010);

\bibitem{Henkel} N. Henkel, R. Nath and T. Pohl, Phys. Rev. Lett. \textbf{104}, 195302 (2010).

\bibitem{SoftSpheres} S. Saccani, S. Moroni, and M. Boninsegni, Phys. Rev. B \textbf{83}, 092506 (2011); S. Rossotti, M. Teruzzi, D. Pini, D. E. Galli, and G. Bertaina, Phys. Rev. Lett. \textbf{119}, 215301 (2017). 

\bibitem{Andreev2} For dipolar condensates a quasi-1D droplet crystal ("fragmented-condensate solid") was predicted in S. V. Andreev, Phys. Rev. Lett. \textbf{110}, 146401 (2013).

\bibitem{Ronen} S. Ronen, J. Phys. B: At. Mol. Opt. Phys. \textbf{42}, 055301 (2009).

\bibitem{Zwerger} J. Hofmann and W. Zwerger, Phys. Rev. X \textbf{7}, 011022 (2017).

\bibitem{Pekar} L. D. Landau, Phys. Z. Sowjetunion \textbf{3}, 644 (1933); S. I. Pekar, Zh. Eksp. Teor. Fiz. \textbf{16}, 335 (1946).

\bibitem{Timmermans1} F. M. Cucchietti and E. Timmermans, Phys. Rev. Lett. \textbf{96}, 210401 (2006).

\bibitem{Experiments} R. Lopes, C. Eigen, A. Barker, K. G. H. Viebahn, M. Robert-de-Saint-Vincent, N. Navon, Z. Hadzibabic, and R. P. Smith, Phys. Rev. Lett. \textbf{118}, 210401 (2017).

\bibitem{Feynman2} R.P. Feynman and M. Cohen, Phys. Rev. \textbf{102}, 1189 (1956).

\bibitem{Miller} A. Miller, D. Pines, and P. Nozieres, Phys. Rev. \textbf{127}, 1452 (1962).

\bibitem{Pines1} P. Nozieres and D. Pines, \textit{Theory of Quantum Liquids, Vol. II:
Superfluid Bose Liquids} (Addison-Wesley, Redwood City, Calif., 1964, 1990).

\bibitem{Taylor} J. R. Taylor, \textit{Scattering Theory} (John Wiley \& Sons, New York, 1972).

\bibitem{Bruun1} R. S. Christensen, J. Levinsen, and G. M. Bruun, Phys. Rev. Lett. \textbf{115}, 160401 (2015).

\bibitem{Kim} Joon Hyun Kim, Deokhwa Hong, and Y. Shin, arXiv:1907.10289v1 (2019).

\bibitem{Timmermans2} D. H. Santamore and E. Timmermans, New Journal of Physics \textbf{13}, 103029 (2011).

\bibitem{SI} See Supplemental Information for the details. 

\bibitem{Ticknor} C. Ticknor, C. A. Regal, D. S. Jin, and J. L. Bohn, Phys. Rev. A \textbf{69}, 042712 (2004).

\bibitem{Papp} S. B. Papp, J. M. Pino, and C. E. Wieman, Phys. Rev. Lett. \textbf{101}, 040402 (2008).

\bibitem{Klawun} M. Klawunn, A. Pikovski, and L. Santos, Phys. Rev. A \textbf{82}, 044701 (2010).

\bibitem{SantosExcitons} Colin Hubert, Yifat Baruchi, Yotam Mazuz-Harpaz, Kobi Cohen, Klaus Biermann, Mikhail Lemeshko, Ken West, Loren Pfeiffer, Ronen Rapaport, and Paulo Santos, Phys. Rev. X \textbf{9}, 021026 (2019).

\bibitem{footnote} This property suggests further consideration of the magnetoroton from the perspective of the \textit{angulon} theory, see Richard Schmidt and Mikhail Lemeshko, Phys. Rev. Lett. \textbf{114}, 203001 (2015); Enderalp Yakaboylu, Bikashkali Midya, Andreas Deuchert, Nikolai Leopold, and Mikhail Lemeshko, Phys. Rev. B \textbf{98}, 224506 (2018).      

 
\end{thebibliography}
\end{document}